\documentclass[11pt]{article}
\usepackage{graphicx}
\usepackage{amsmath}

\begin{document}
\title{Band Mapping in One-Step Photoemission Theory:
Multi-Bloch-Wave Structure of Final States and 
Interference Effects}

\author{E.E. Krasovskii,$^1$
% \\ 
% {\footnotesize\it Institut f\"ur Theoretische Physik, 
% Christian-Albrechts-Universit\"at, 
% Leibnizstrasse 15, 24098 Kiel, 
% Germany} 
\and 
V.N. Strocov,$^2$
% \\ 
% {\footnotesize\it Swiss Light Source, Paul Scherrer Institute, 
% CH-5232 Villigen PSI, Switzerland} 
\and
N. Barrett,$^3$
% \\ 
% {\footnotesize\it CEA-DSM/DRECAM-SPCSI, CEA-Saclay, 
% 91191 Gif-sur-Yvette, France}
\and 
H.~Berger,$^4$
% \\ 
% {\footnotesize\it Institut de Physique Appliquee, EPFL, CH-1015 Lausanne, Switzerland} 
\and    
W.~Schattke,$^{1,5}$
% \\ {\footnotesize\it
% Institut f\"ur Theoretische Physik, 
% Christian-Albrechts-Universit\"at, Leibnizstrasse 15, 24098 Kiel, 
% Germany,}
% \\
% {\footnotesize\it
% Donostia International Physics Center, 
% P.$^{\circ}$ Manuel de Lardiz{\'a}bal, 4, 20018 Donostia-San Sebasti{\'a}n, 
% Spain}
\and 
R. Claessen$^6$\\
\
% \\ {\footnotesize\it
% Experimentelle Physik 4, Universit\"at W\"urzburg, 
% 97074 W\"urzburg, Germany}
\and
{\footnotesize\it $^1$Institut f\"ur Theoretische Physik, 
Christian-Albrechts-Universit\"at, 
Leibnizstrasse 15, 24098 Kiel, 
Germany} 
\and 
{\footnotesize\it $^2$Swiss Light Source, Paul Scherrer Institute, 
CH-5232 Villigen PSI, Switzerland} 
\and
{\footnotesize\it $^3$CEA-DSM/DRECAM-SPCSI, CEA-Saclay, 
91191 Gif-sur-Yvette, France}
\and 
{\footnotesize\it $^4$Institut de Physique Appliquee, EPFL, CH-1015 Lausanne, Switzerland} 
\and    
{\footnotesize\it
$^5$Donostia International Physics Center, 
P.$^{\circ}$ Manuel de Lardiz{\'a}bal, 4, 20018 Donostia-San Sebasti{\'a}n, 
Spain}
\and
{\footnotesize\it
$^6$Experimentelle Physik 4, Universit\"at W\"urzburg, 
97074 W\"urzburg, Germany}
}

%}
%\affiliation{
% {
% $^1$Institut f\"ur Theoretische Physik, 
% Christian-Albrechts-Universit\"at, Leibnizstrasse 15, 24098 Kiel, 
% Germany\\
%
% $^2$Swiss Light Source, Paul Scherrer Institute, 
% CH-5232 Villigen PSI, Switzerland\\
%
% $^3$CEA-DSM/DRECAM-SPCSI, CEA-Saclay, 91191 Gif-sur-Yvette, France\\
%
% $^4$Institut de Physique Appliquee, EPFL, CH-1015 Lausanne, Switzerland\\
%
% $^5$Donostia International Physics Center, 
% P.$^{\circ}$ Manuel de Lardiz{\'a}bal, 4, 20018 Donostia-San Sebasti{\'a}n, 
% Spain\\
%
% $^6$Experimentelle Physik 4, Universit\"at W\"urzburg, 
% 97074 W\"urzburg, Germany}

% \begin{center}
% \bf Band Mapping in One-Step Photoemission Theory:\\
% Multi-Bloch-Wave Structure of Final States and 
% Interference Effects
% \end{center}

% \centerline{\small E.E. Krasovskii$^1$, V.N. Strocov$^2$, 
% N. Barrett$^3$, H.~Berger$^4$,    
% W.~Schattke$^{1,5}$, and R. Claessen$^6$}

% \medskip

% {
% \noindent
% \footnotesize\it
% $^1$Institut f\"ur Theoretische Physik, 
% Christian-Albrechts-Universit\"at, Leibnizstrasse 15, 24098 Kiel, 
% Germany\\
%
% $^2$Swiss Light Source, Paul Scherrer Institute, 
% CH-5232 Villigen PSI, Switzerland\\
%
% $^3$CEA-DSM/DRECAM-SPCSI, CEA-Saclay, 91191 Gif-sur-Yvette, France\\
%
% $^4$Institut de Physique Appliquee, EPFL, CH-1015 Lausanne, Switzerland\\
%
% $^5$Donostia International Physics Center, 
% P.$^{\circ}$ Manuel de Lardiz{\'a}bal, 4, 20018 Donostia-San Sebasti{\'a}n, 
% Spain\\
%
% $^6$Experimentelle Physik 4, Universit\"at W\"urzburg, 
% 97074 W\"urzburg, Germany}\\

\maketitle 

\begin{abstract}
%%%%%%%%%%%%%%%%%%%%%%%%%%%%%%%%%%%%%%%%%%%%%%%%%%%%%%%%%%%%
A novel Bloch-waves based one-step theory of photoemission is 
developed within the augmented plane wave formalism. Implications 
of multi-Bloch-wave structure of photoelectron final states for
band mapping are established. Interference between Bloch components 
of initial and final states leads to prominent spectral features
with characteristic frequency dispersion experimentally observed
in VSe$_2$ and TiTe$_2$. Interference effects together with a 
non-free-electron nature of final states strongly limit the 
applicability of the common direct transitions band mapping approach,
making the tool of one-step analysis indispensable.
%%%%%%%%%%%%%%%%%%%%%%%%%%%%%%%%%%%%%%%%%%%%%%%%%%%%%%%%%%%%
\end{abstract} 

%{79.60.-i, 71.20.-b, 71.15.-m}

Angle resolved photoemission (ARPES) is the most powerful tool to study 
electronic structure of crystals \cite{schattke_book}. The basis for the 
interpretation of experimental results is provided by the one-step theory 
of photoemission \cite{FeibelEast74}, in which the photoelectron initial 
states are one-particle eigenstates of the semi-infinite crystal, and the 
final state is the time reversed LEED state $|\Phi\rangle$. The interaction 
with the many-electron system of the crystal gives rise to inelastic
scattering and leads to a finite mean free path of the photoelectron.
In spite of the lacking translational invariance in the surface perpendicular 
direction the concept of 3-dimensional band structure has proved very fruitful
for interpretation of ARPES spectra \cite{STR98}. In the band structure picture
the finite photoelectron mean free path is seen as a crystal momentum 
uncertainty of the final state Bloch waves, whose immediate effect is to 
broaden the spectral structures due to direct transitions. For a reasonably 
small momentum broadening it is often assumed that the bulk photoemission 
can be understood in terms of the $E(k_\perp)$ dispersion of occupied and 
unoccupied bands.

The one-step model, however, implies a more complicated picture: following 
Slater \cite{Slater37} and Pendry \cite{Pendry76} the outgoing electron is 
described by a pure state represented by {\em a damped wave function}: the 
inelastic scattering is taken into account by adding an imaginary part 
$-iV_{\rm i}$ to the potential in the crystal half-space, so that 
$|\Phi\rangle$ is an eigenfunction of a non-Hermitean Hamiltonian 
corresponding to a real eigenvalue $E_{\rm f}$ -- the photoelectron final 
energy. The imaginary term causes the LEED function to decay in space and 
provides the model with the desired surface sensitivity. The final state,
thus, retains its fully coherent nature, which governs the interference
of the transition matrix elements between the Bloch constituents of the 
final and the initial state.

\begin{figure}[h]    %%%%%%%%%%%%%%%%%%%%%%%%%%%%%%%%%%%%%%%%%%%%%%%%%% FIG 1
\begin{center}
\includegraphics[width=0.85\textwidth, bb= 10 10 800 800]{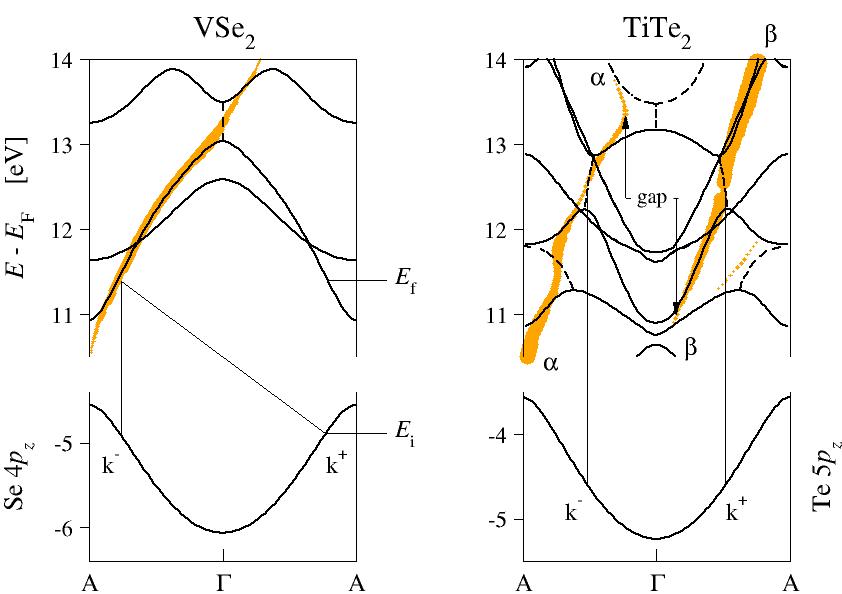}
\caption{\label{one-step} 
Se 4$p_z$ and Te 5$p_z$ bands of VSe$_2$ and TiTe$_2$ and fragments of 
unoccupied band structure. Full curves display real band structure and
dashed curves real lines of complex band structure with $V_{\rm i}=0$.
Thick lines show conducting constituents of the final states calculated 
with optical potential $V_{\rm i}=1$~eV. The line thickness is proportional to
the current carried (absorbed) by this partial wave \cite{absorbed_current}. 
In TiTe$_2$ two conducting branches are seen: $\alpha$ and $\beta$. Straight 
lines illustrate cases of indirect (left panel) and direct (right panel) 
interference. Arrows in the right panel indicate the final state $k_\perp$ 
gap responsible for the dispersion turning point in TiTe$_2$ at 
$E_{\rm f}=12.5$~eV in Fig.\ref{band_mapping}.
}
\end{center}
\end{figure}

In the simplest case the final state comprises only one dominant Bloch 
wave, as in VSe$_2$ in the energy range shown in Fig.~\ref{one-step}, 
but there may be two or more Bloch constituents, as in TiTe$_2$ in 
Fig.~\ref{one-step}. The photoelectron initial state is a standing wave, 
and in the depth of the crystal it is a 
superposition of the Bloch wave $|\mathbf{k}^+\rangle$ incident from the 
interior of the crystal on the surface and a number of reflected waves 
traveling in the opposite direction. In the simplest case there is only 
one reflected wave $|\mathbf{k}^-\rangle$, see Fig.~\ref{one-step}. 
The dipole matrix elements 
$\langle\Phi | \hat {\mathbf p} | {\mathbf k}^+\rangle$ and
$\langle\Phi | \hat {\mathbf p} | {\mathbf k}^-\rangle$ may 
cancel or enhance each other depending on their phases. Let us consider 
a single-Bloch-wave final state. When the absorbing potential $V_{\rm i}$ 
is small the phases of the matrix elements are unimportant because only 
the matrix element of the momentum conserving transition is large (see 
the transition at $\mathbf{k}^-$ in the left panel of Fig.~\ref{one-step}). 
For realistic values of $V_{\rm i}$, which are of the order of 1~eV, the 
matrix elements become comparable, and the interference of the waves 
becomes noticeable. For brevity, we shall refer to this as {\em indirect} 
interference; its importance increases with inelastic scattering strength.
If the final state is composed of several Bloch states strong interference 
may occur with a negligible momentum broadening, as in TiTe$_2$ in 
Fig.~\ref{one-step}. This we shall call {\em direct} interference. 
Although a composite structure of final states has been noticed earlier 
\cite{composite}, the effect of interference and its implications for the 
band mapping have not been realized. 

                          %%%%%%%%%%%%%%%%%%%%%%%%%%%%%%%%% FIG 1
In the present paper we elucidate the influence of the Bloch wave interference 
on the observed spectra with the aim to understand the connection between a 
photoemission spectrum and the underlying band structure in more detail.
In view of the phenomenological nature of the final state the question arises 
whether the interference effects are correctly treated in the one-step theory
and whether the multi-Bloch-wave structure of the final state may provide
additional information about electron states. To clarify these issues, we have 
performed a combined theoretical and experimental study of normal emission 
from VSe$_2$ and TiTe$_2$ layered crystals.

Our theory is based on the band structure approach both to final and 
to initial states \cite{KRAS04_A}. The LEED state is a solution of a 
scattering problem for a plane wave incident from vacuum. In the bulk 
of the crystal the wave function is given in terms of the complex band 
structure, and it is continued into the surface region by solving a 
Cauchy problem. A detailed description of the methodology for an 
all-electron potential of general shape within the augmented plane wave 
(APW) formalism has been presented in Ref.~\cite{KRAS04_B}. A novel 
aspect of the present method is that the same scattering technique is 
used for initial states, only the incident wave is now the Bloch wave 
$|\mathbf{k}^+\rangle$ \cite{comment1}.

The APW based Bloch wave scattering theory offers an accurate and 
efficient computational scheme, whose characteristic features are: 
(i)   all multiple scattering in the crystal and
      at the surface is taken into account within a full-potential technique;
(ii)  connection between the spectra and the complex band structure    
      of the crystal is straightforward and transparent;
(iii) no electron absorption is introduced in calculating initial states.
The treatment of initial states is the main difference between the
present theory and more traditional KKR-based scattering theories: 
the layer doubling method of Ref.~\cite{Pendry76} essentially relies 
on the imaginary term in the Hamiltonian, and in real-space techniques 
\cite{woods01} the Bloch character of the electron states is obscured.  
 
The ARPES experiment was performed at the SuperACO synchrotron in LURE, 
France. The spectra were measured at normal emission with photon energies 
between 11.5 and 35 eV. The synchrotron radiation polarization vector was 
set at an angle of 45$^o$ to the surface normal in the $M\Gamma M'$ azimuth. 
The combined monochromator and analyzer energy resolution varied from 23 to 
130 meV with increase of the photon energies through the experimental range.

\begin{figure}[htb]   %%%%%%%%%%%%%%%%%%%%%%%%%%%%%%%%%%%%%%%%%%%%%%%%% FIG 2
\begin{center}
\includegraphics[width=0.85\textwidth, bb= 10 10 800 800]{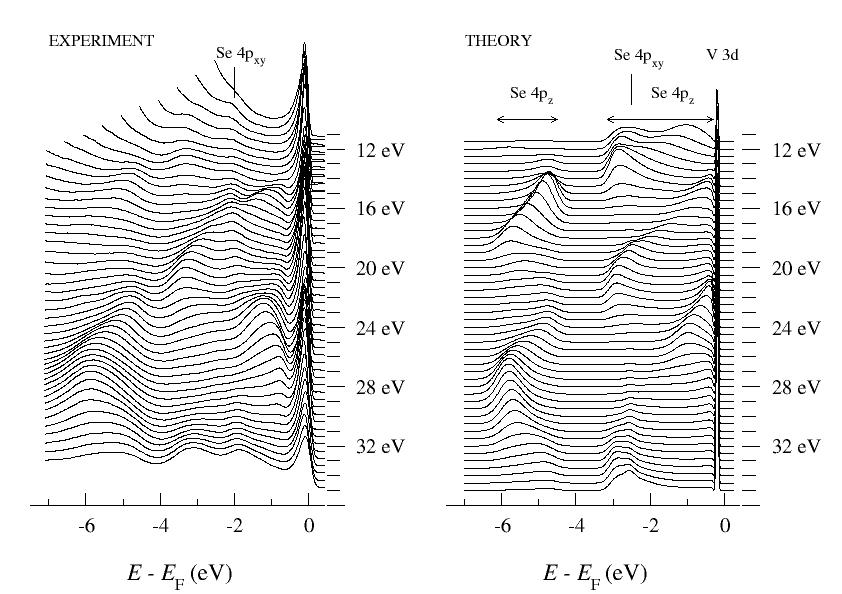}
\caption{\label{Fig2} 
Normal emission energy distribution curves of VSe$_2$ parametrized by photon
energy: experiment (left) and theory (right). Spectra are normalized to the 
same integral intensity.
}
\end{center}
\end{figure}                          %%%%%%%%%%%%%%%%%%%%%%%%%%%%%%%%% FIG 2

\begin{figure}       %%%%%%%%%%%%%%%%%%%%%%%%%%%%%%%%%%%%%%%%%%%%%%%%%% FIG 3
\begin{center}
\includegraphics[width=0.85\textwidth, bb= 10 10 800 800]{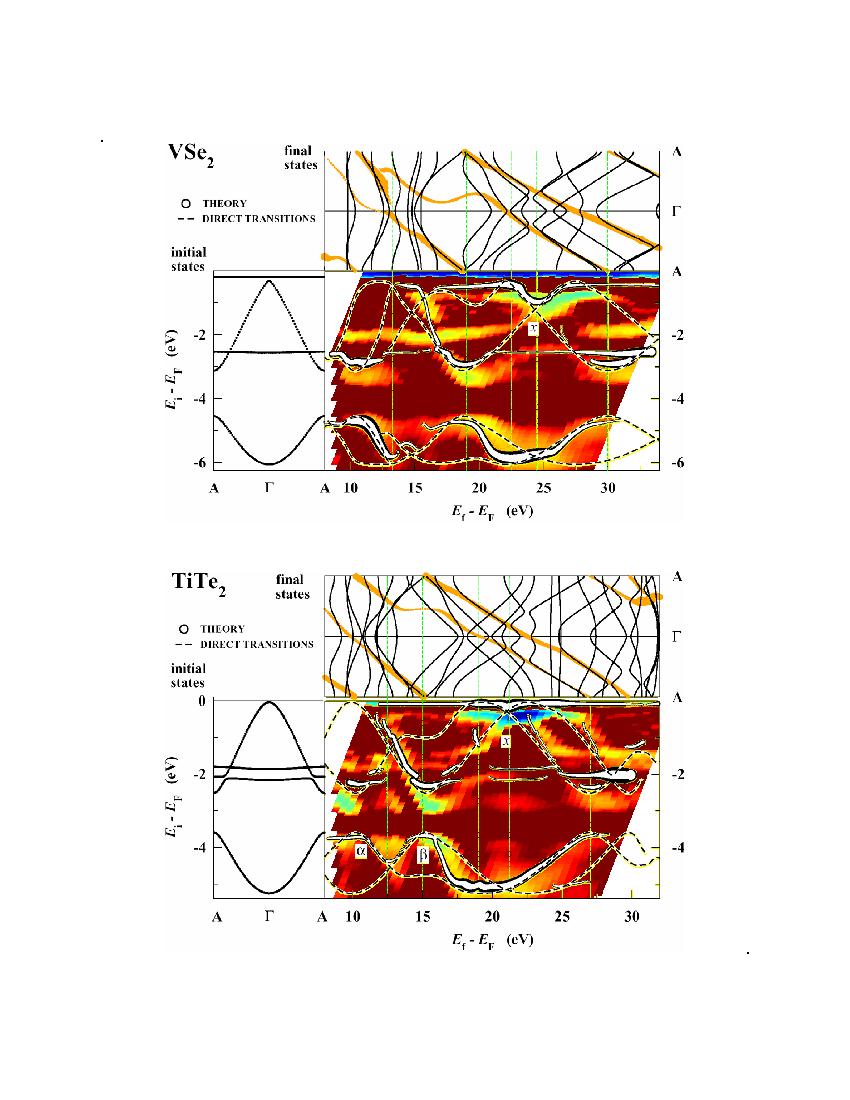}
\caption{\label{band_mapping} 
(color online) Peak dispersion diagrams for normal emission from VSe$_2$ and 
TiTe$_2$. Calculated band structure for initial states is shown in left panels
and for final states in upper panels with thin lines. In the final states 
panels thick lines show the branches of complex band structure (CBS) that
most strongly contribute to the final state (thickness is proportional to 
the current absorbed by the Bloch component of the LEED state). In the 
dispersion diagram dashed lines show the location of direct transitions 
to the conducting branches of CBS (lines due to nondispersive $p_{xy}$ and 
V $3d$ states are not shown). The peak locations in the theoretical EDCs is 
shown by white lines with black fringes (thickness is proportional to peak 
intensity). They are superimposed on the experimental second derivative map 
$-d^2I(E_{\rm i})/dE_{\rm i}^2$ shown in logarithmic colorscale (areas with 
$-d^2I(E_{\rm i})/dE_{\rm i}^2<0$ are clipped).
}
\end{center}
\end{figure}                          %%%%%%%%%%%%%%%%%%%%%%%%%%%%%%%%% FIG 3

Experimental and theoretical spectra for VSe$_2$ are compared in
Fig.~\ref{Fig2}. Our theory correctly reproduces all the 
main structures and their dispersion with photon energy. Similar 
level of agreement is obtained for TiTe$_2$. Final states were validated 
by comparing the energy dependence of the current carried by the LEED state
with the electron transmission measurements \cite{fstite}.  A detailed 
comparison of the peak dispersion in theory and experiment for VSe$_2$ 
and  TiTe$_2$ is presented in Fig.~\ref{band_mapping}. Well visible are 
the discrepancies due to simplified treatment of quasiparticles within 
the local density approximation (LDA): the calculated upper Se 4$p_z$ band 
is shifted upwards by 0.3~eV, and the Se 4$p_{xy}$ band downwards by 
0.5~eV \cite{disagreement}. Discrepancies of the same character and size 
are observed for the Te 5$p$ states in TiTe$_2$. In both crystals, below 
$E_{\rm f}\sim 16$~eV the measured final energy location of the peaks is 
well reproduced, and above 16~eV the measured dispersion curves are stretched 
towards higher energies. It is interesting that although the unoccupied bands 
of VSe$_2$ are shifted by 3 to 4~eV upwards relative to the bands of TiTe$_2$ 
the self-energy effects are seen to start growing at roughly the same energy.

The unoccupied bands look rather complicated, but the calculation points 
out the few waves important for the electron's escape to vacuum. Their 
structure is relatively simple and transparent (in Fig.~\ref{band_mapping} 
the conducting bands are shown by thick lines in the final-states graphs). 
We note, however, that a nearly free electron model is inapplicable here 
simply because over wide energy regions there are two waves strongly
contributing to the LEED function. The Bloch-wave scattering theory enables 
us to explicitly separate out the effect of $k_\perp$ dispersion of the 
initial and final Bloch waves and to reveal the more subtle factors due 
to the structure of the wave functions. It is instructive to plot the direct 
transitions lines -- a procedure frequently used in comparing experimental 
spectra with a calculated band structure: the dashed lines in 
Fig.~\ref{band_mapping} show the points with coordinates 
$[E_{\rm f}(k_\perp),E_{\rm i}(k_\perp)]$, with $k_\perp$ running over the 1D 
Brillouin zone. The calculated peak dispersion is seen to often deviate from 
the dashed lines. A vivid example of such an intrinsic shift 
\cite{intrinsic_shift} is the downward dispersion of the lower $p_z$ bands 
between 20 and 21~eV in VSe$_2$ and between 16 and 17~eV in TiTe$_2$. The 
analysis shows that it is not just an overlap of two transitions but their 
indirect interference that determines the dispersion of the peak.
 
A clear manifestation of the two-branch structure of the LEED states is
a curious dispersion of the lower Te 5$p_z$ peak between $E_{\rm f}=11$ 
and 15~eV: it disperses downwards but does not reach the bottom of the 
band and vanishes at $\sim 13$~eV to appear again at the valence band 
maximum less than 3~eV higher in energy. This happens because the leading 
final state wave hops from branch $\alpha$ to branch $\beta$ 
(see Fig.~\ref{one-step}) at $E_{\rm f}\approx 12$~eV. In VSe$_2$ 
at low energies only one Bloch wave dominates (see Fig.~\ref{one-step}), 
which allows us to more clearly see the effect: in VSe$_2$ the peak disperses 
almost to the band minimum, and the maximum is reached some 6~eV higher 
in energy. Although in TiTe$_2$ the direct interference influences the 
shape of the spectra, it is not essential for understanding this behavior: 
the rapid change from minimum to maximum is caused by the $k_\perp$ gap 
between the $\alpha$ and $\beta$ branches. 

\begin{figure}[h]    %%%%%%%%%%%%%%%%%%%%%%%%%%%%%%%%%%%%%%%%%%%%%%%%%% FIG 4
\begin{center}
\includegraphics[width=0.8\textwidth, bb= 10 10 800 800]{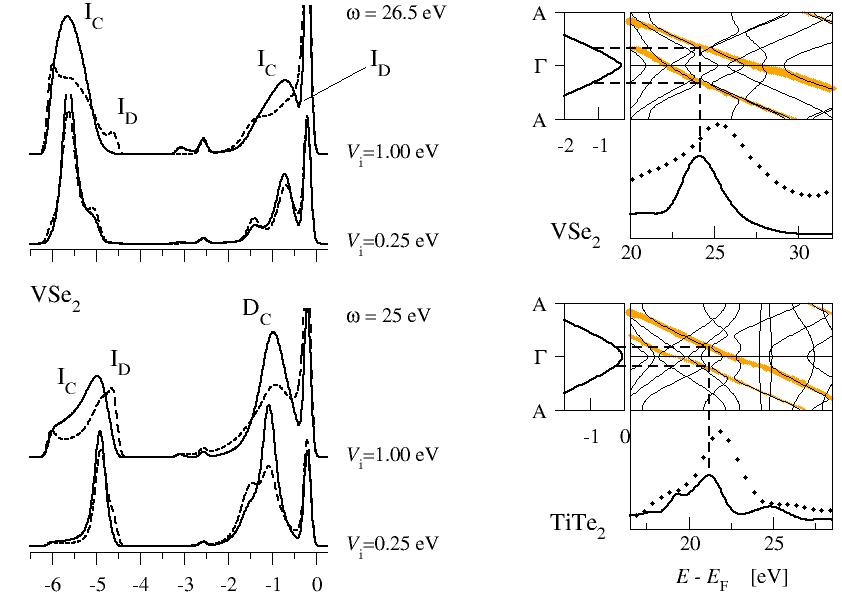}
\caption{
\label{coherence} 
  Interference effects in VSe$_2$. Left: Full lines are energy distribution 
  curves calculated within the one-step theory, and dashed lines the result 
  of incoherent summation of the emission from the $|{\mathbf k}^-\rangle$ 
  and $|{\mathbf k}^+\rangle$ states. Structures affected by indirect (I)
  and direct (D) interference are marked by symbols I$_{\rm C,D}$ and 
  D$_{\rm C}$, with subscripts referring to constructive (C) and destructive 
  (D) interference. Direct interference is recognized by its influence on the 
  spectra with $V_{\rm i}=0.25$~eV. Right: Initial and final states whose 
  direct interference causes the feature $x$ in the dispersion diagrams in 
  Fig.~{\protect\ref{band_mapping}}. Lower panels show calculated (lines)
  and measured (circles) constant initial state spectra for the energies
  $E_{\rm i}$ at which the strongest intensity enhancement is observed. 
}
\end{center}
\end{figure}

A constructive direct interference plays an important role in formation 
of the structures $x$ in VSe$_2$, around $E_{\rm f}=24.5$~eV at 
$E_{\rm i}\approx-1.2$~eV, and in TiTe$_2$, around $E_{\rm f}=21.5$~eV at 
$E_{\rm i}\approx-0.5$~eV, see Fig.~\ref{band_mapping}. In both
materials the experimentally observed intensity enhancement correlates
well with the crossing points of the two direct transition lines.
In VSe$_2$ the structure appears well below the valence band maximum, and
the wave vectors of both final state constituents are well distant from 
the $\Gamma$ point. In TiTe$_2$ the dispersion is less pronounced, which
is consistent with the differences in the band structures (see right panels 
of Fig.~\ref{coherence}): in TiTe$_2$ the crossing occurs much closer to the 
$\Gamma$ point than in VSe$_2$, and at the same time the initial state 
dispersion is weaker.

The interference origin of the features is illustrated by Fig.~\ref{coherence},
where we compare VSe$_2$ spectra for $\hbar\omega=25$ and 26.5~eV calculated 
within the one-step theory with the spectra resulting from incoherent 
summation of the squared moduli of the elements 
$\langle\Phi|\hat{\mathbf p}|{\mathbf k}^+\rangle$ and  
$\langle\Phi|\hat{\mathbf p}|{\mathbf k}^-\rangle$. For a small momentum 
broadening, $V_{\rm i}=0.25$~eV, the deviation of a coherent curve from its 
incoherent counterpart reveals the effect of direct interference. The 
indirect interference leads to differences that are seen in $V_{\rm i}=1$~eV 
curves and not in $V_{\rm i}=0.25$~eV ones. Our important finding is that 
the structures $x$ are due to a strong final-state energy dependence of the 
direct interference: in Fig.~\ref{coherence} the intensity enhancement 
D$_{\rm C}$ is large for $\hbar\omega=25$~eV and negligible for 
$\hbar\omega=26.5$~eV. 
                          %%%%%%%%%%%%%%%%%%%%%%%%%%%%%%%%% FIG 4

To illustrate the importance of the interference structures for unambiguous 
band mapping, we show in the right panels of Fig.~\ref{coherence} constant 
initial state spectra for the energies $E_{\rm i}$ at which the strongest 
intensity enhancement is observed. For VSe$_2$ the calculated and measured 
initial energies are $E_{\rm i}=-1$ and $-1.2$~eV, respectively (see 
Fig.~\ref{Fig2}), and for TiTe$_2$ it is $-0.2$ and $-0.5$~eV.
The final state energy of the direct interference transition is seen 
to be reliably determined in experiment (as expected, the experimental 
structures are shifted to higher $E_{\rm f}$ energies due to self-energy 
effects.) Because the $x$ structure corresponds to a symmetric $k_\perp$ 
location of the two final state branches the $k_\perp$ assignment of the
initial state is stable to the self-energy shift of the final states -- 
in contrast to the case of a single-Bloch-wave final state.

To summarize, a Bloch-waves based theory of photoemission is successfully 
applied to trace back the experimentally observed peak dispersion in VSe$_2$
and TiTe$_2$ to the band structure of initial and final states. Owing to 
Bloch wave interference, information on just dispersion of initial and 
conducting final states is often insufficient to understand the experimental 
results; naive band mapping may be misleading if not supported by accurate 
calculations of intensities. At the same time, our results indicate a dominant
role of direct transitions, which enables reliable band mapping in the absence
of interference. 

For the first time spectral features are identified that arise from the 
multi-Bloch-wave structure of the final states. Structures of this novel
type are observed as turning points in the peak dispersion diagrams that 
correspond neither to minima nor to maxima of the initial state bands. 
Owing to the dominant role of direct transitions and the interference 
enhancement of the emission intensity, such points provide information 
about the wave vectors of the final state constituent and allow a more 
detailed band mapping.

The deviation of the calculated peak dispersion from the direct transition 
lines is the result of the Bloch wave interference in the initial states.
This effect is intimately connected to the optical-potential approach
to inelastic scattering in the one-step theory. The good agreement with 
experiment in cases of strong interference, thus, provides the most direct 
proof so far in favor of the damped waves treatment of inelastic scattering.

The work was supported by Deutsche Forschungsgemeinschaft via
Forschergruppe FOR 353 and grant CL 124/5-2, and the experiments 
at LURE by the EC within the Access to Research Infrastructure program.


\begin{thebibliography}{12}
\bibitem{comment1} The present theory is scalar-relativistic, which is quite 
                   acceptable for VSe$_2$, but is a considerable approximation
                   for Te 5$p$ states in TiTe$_2$. In the latter case we used 
                   relativistic initial state energies and one-component 
                   scalar-relativistic wave functions.
\end{thebibliography}
\end{document}